ARTICLE

# How Do Water Filled Traffic Barriers Shake a Suspension Bridge?

Guanni Qu[1,#], Tianai Yue[2,#,*], Xiaoyu Zhang[3] and Shibiao Wei[3]

[1]The Shipley School, Bryn Mawr, USA
[2]Tianjin Nankai High School, Tianjin, China
[3]Xi'an Gaoxin No. 1 High School, Xi'an, China
*Corresponding Author: Tianai Yue. Email: yuetianai1018@sina.com
#These authors contributed equally to this work

## ABSTRACT

The present study stems from the realization that the general problem relating to the analysis of wind-induced vibrations in suspension bridges still requires significant attention. Sidewalk railings, overhaul tracks, and deflectors are known to largely affect such dynamics. Here, the influence of a row of water-filled traffic barriers on the response of a sample suspension bridge is investigated numerically. It is shown that the existence of water barriers causes flow separation and non-negligible vortices with respect to the condition with no water barriers. The vortex shedding frequency at the far end is around 41.30 Hz, relatively close to the real vibration frequency. It is also shown how different incoming angles of attack can change the flow field around the bridge cross-section and the vortex detachment frequency.

## 1 Introduction

A long suspension bridge may undergo abnormal vibration when the wind speed near the bridge is far less than the upper limit. One reason for this is that, during bridge maintenance, the temporary placement of water barriers at the roadside damages the aerodynamic shape of the bridge's main girder by vortex vibration. A water barrier is a plastic, water-filled movable barrier that is used to divide the road. The water barrier is hollow inside and has a hole at the top to increase the weight. Additionally, the water barrier is shaped like a saddle, that is, narrow at the top and wide at the bottom.

Vortex vibration is a resonance phenomenon that occurs when the vortex shedding frequency is close to the structure's natural frequency [1,2], and has both the characteristics of self-excited vibration and forced vibration. Although vortex-induced vibration generally belongs to limited amplitude vibration, high-frequency and long-time vibration will cause fatigue damage to the bridge structure. Additionally, an excessively large amplitude will affect the driving safety on the bridge. The water-filled traffic barriers may be a potential cause for the vortex vibration of bridges. Therefore, it is necessary to increase awareness with regard to the potential risks of adding objects to bridges.

Because the aerodynamic shape directly determines the characteristics of flow separation and vortex formation, the aerodynamic shape of the main girder is very important for the occurrence and amplitude of vortex vibration [3,4]. The wide-body flat steel box girder of the suspended bridge is a type of near-streamline section, and its aerodynamic characteristics are obviously different such others as the blunt sections, truss girder, open box girders [5–8]. Thus, even a small change in the shape may lead to an obvious quantitative or even qualitative change in the vortex response [4]. Therefore, it is important to investigate the flow separation and vortex vibration phenomenon of long-span suspension bridges and the influence of structures attached to the bridges. In such abnormal vibration incidents, a row of water-filled traffic barriers is placed on the bridge deck to separate the road surface. This change may have a certain effect on the aerodynamic shape of the main girder of the bridge, which in turn causes vortex vibration. Therefore, an in-depth analysis of the placement of the water barriers on the frequency of vortex shedding will help our understanding of the cause of the abnormal vibration of the suspension bridge.

This study analyzed the influence of placing water barriers on the vortex vibration of a suspended bridge. The placement of a water barrier at the distal end of the bridge induces a vortex shedding frequency, which is very similar to the resonance frequency of the bridge, and vastly increases the vertical vortex-bending response of the bridge. In the simulations with negative incoming angles of attack (–5°), the circumferential vortex flow field and wake vortex detachment frequency were similar to the resonance frequency of the investigated suspended bridge.

## 2 Related Work
### 2.1 Long-span Bridge

As the span of a suspension bridge increases, its structure becomes lighter, less damped, and less rigid. Therefore, the suspension bridge becomes sensitive to wind and vibration can easily be produced. From 1818 to 1940, at least 11 suspension bridges were destroyed by wind [9]. In 1940, the Tacoma Narrows Bridge (853 m) vibrated strongly at a wind speed lower than 20.0 m/s and the main girder was broken. This wind-induced bridge damage has paved the way for research on wind-induced vibration and the development of the aeroelasticity theory of long-span bridges. There are currently 22 suspension bridges with a span of more than 1,000 m, out of which 12 are made of steel box girders, and 10 are made of steel truss girders [10,11]. During design and operation, seven of these bridges have had problems caused by wind-induced vibration; six of them underwent flutter and one underwent vortex vibration. The flat box girder first appeared on Seven Bridge in England in 1966. The new bridge section's wind resistance coefficient and steel consumption are much smaller than those of the truss girder. The flutter performance of the new bridge section is satisfactory and reduces the maintenance workload. Subsequently, this near-streamline section was widely used in the design and construction of long-span bridges and has been in use until now.

Owing to their low stiffness and damping, these long-span bridges are very sensitive to wind. Therefore, the vortex-induced resonance phenomenon has been increasingly observed in actual operations. With a main span of 1624 m, the Danish Great Belt Bridge underwent multiple vertical bending vortex vibrations with an amplitude of 30–50 cm at wind speeds of 5–10 m/s during the construction phase, and multiple vortex vibration accidents have occurred during its operation. Under wind speeds of 5–10 m/s during the construction stage, the main span of the 1624-m Danish Great Belt Bridge exhibited multi-order vertical bending vortex vibration with an amplitude of 30–50 cm. Additionally, several vortex vibration accidents have occurred during its operation. In the Osteroy Bridge in Norway, $1^{st}$–$5^{th}$ order vertical bending vortex-induced resonance has occurred at a wind speed of only 5–8 m/s and amplitude of 50 cm. For the cross-sea Xihoumen Bridge, whose main span is 1650 m, multi-order vertical bending vortex vibration with an amplitude of up to 16–18 cm has occurred at wind speeds of 6–11 m/s. In such cases, the bridge is closed and traffic is not allowed.

## 2.2 Research on Vortex-Induced Vibration

Vortex-induced vibration at low wind speed has increasingly attracted attention from researchers [4], particularly with regard to flexible kilometer-span bridges. The initial wind speed of the vortex-induced vibration is mostly normal (for example, 8 m/s). There are three main types of vortex vibration: vertical-bending vortex vibration, torsional vortex vibration, and bending-torsion coupling vibration. In the beginning, the resonance caused by vortex shedding is forced vibration. In the later period of vibration, the motion of the structure begins to affect the vortex shedding and therefore has the characteristics of self-excitation. Presently, the mechanism of vortex-induced vibration is not clear. In recent decades, Matteoni, Bearman, and Sarpkaya reviewed the research on vortex-induced vibration [12–14]. Moreover, many studies have deduced the vortex-induced vibration of structures and reached reasonable conclusions, which provide a basis for further research.

Various studies have investigated the phenomenon of vortex-induced vibration in structures. The methods for investigating bridge vortex-induced vibration mainly include theoretical analysis, numerical methods, and wind tunnel tests. With the rapid development of computer technology, computational fluid dynamics (CFD) approaches have played an increasingly important role. Mittal et al. [15] used the Finite Element Method to numerically simulate the flow around a column and cylinder under a low Reynolds number, and observed a 2S- or 2P-shaped wake vortex at different stages in the locked zone. Fang et al. [16] calculated the lateral vortex-induced vibration of a two-dimensional square column under high Reynolds number using their own loosely coupled program, and observed that the vortex shedding mode changed from out-of-phase 2S shape to an in-phase 2S shape. They solved the Navier–Stokes equation using the difference method and the vibration equation using the Newmark-β method, and analyzed the interaction between the cylinder and the fluid using the dynamic grid method. Thus, they observed the beating and locking phenomena of the vortex-induced vibration. Their calculated and experimental results are in good agreement with the results obtained by previous studies [17,18]. The FLUENT RANS method was used to simulate the vortex-induced resonance of the cylinder and the wind-rain-induced vibration of the stay cable [19,20]. Larsen et al. [9] used his developed separated vortex program called DVMFLOW to investigate the wind-induced stability of various bridge sections such as the Tacoma Narrows Bridge and Gibraltar Strait Bridge and summarized the influence of the main girder section on the wind-induced stability of suspension bridges and cable-stayed bridges.

## 2.3 Research of Vortex-Induced Vibration Characteristics of Approximately Streamlined Box Girder

The above-mentioned studies have generally shown that the vortex shedding intensity is large, the damping is small, and the positive angle of attack corresponds to high-amplitude vortex-induced vibration. The vortex-induced vibration of the cross-section is closely related to the vortex shedding pattern in the near-wake region, which causes the formation of vortex vibration. The critical wind speed, amplitude, lock-in area, and wake mode of this type of cross-section will change dramatically with the cross-section shape and effect of the Reynolds number.

Previous studies have investigated the structure of the suspension bridge. This type of bridge is affected by the Earth's self-rotation in the solar and lunar gravitational fields, which gives rise to stress when potential energy is present [21]. Several models have introduced the new structure and construction of the steel deck pavement and analyzed the causes of damage to the pavement [22]. By considering the influence of the section's aerodynamic shape, Guan et al. [23] investigated the vortex vibration performance of the inverted trapezoidal integral steel box girder with or without railings. By measuring the surface pressure of the box girder section, the mechanism through which the railings affect the vortex vibration performance of the bridge section was elucidated. Qin et al. [24] conducted a full bridge model wind tunnel test on a split double-span bridge deck steel box girder and analyzed the mechanism through which the double-span main girder generates two vortex vibration zones. Zhao et al. [25] analyzed the changes in the vortex vibration of the main girder with and without fixed horizontal aerodynamic wings on both sides of the split bluff body double-box steel box girder by conducting a wind tunnel test on a segment model with a scale of 1/50. The effects of sidewalk railings, overhaul tracks, and deflectors on the vortex vibration performance of flat steel box girders have been investigated through large-scale wind tunnel tests on a segment model [26].

The above-mentioned studies have shown that the structures attached to the bridge deck have a significant effect on the vortex-induced vibration characteristics of the section. In practice, to satisfy the operational needs of bridges, more changes should be made to accessories, such as the ventilation condition of the railings, the position of the railings, and the change of the track position of the maintenance vehicle. Presently, there is a lack of relevant studies investigating and confirming the effect of these modifications on the aerodynamic characteristics of the cross-section.

# 3 Research Data

## 3.1 Model of Sample Suspended Bridge

The suspended bridge shown in Fig. 1 was considered to demonstrate the vortex-induced vibration. The main cable span is 312 m + 888 m + 353.484 m, the design speed is 120 km/h, and there are six lanes. The stiffening girder of the bridge is a fully welded, flat, closed, streamlined single-chamber steel box girder with an air nozzle, beam width of 35.6 m, center height of 3.012 m, and net height of 2.99 mm. Because the design is close to streamline and the steel box girder structure has a small wind-facing area and large torsional stiffness, the bridge has high resistance against wind-induced instability. The design wind speed is 8 m/s.

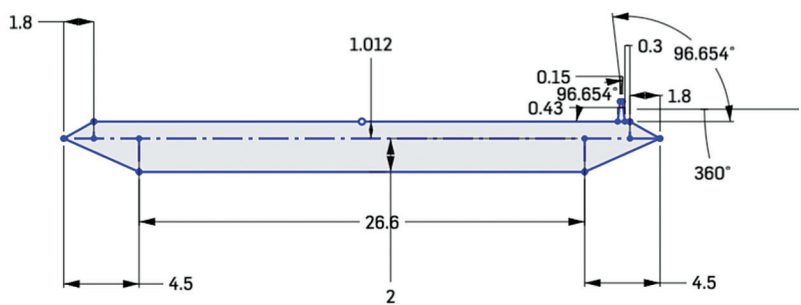

**Figure 1:** Two-dimensional simplified model of suspended bridge

In this study, a geometric cross-section model was used for simulation according to the real geometric dimensions of the investigated suspended bridge. The simplified model is shown in Fig. 1 and the model parameters are listed in Table 1.

**Table 1:** Parameters of simplified two-dimensional model of suspended bridge

| Item | m |
| --- | --- |
| Width of the bottom | 26.6 |
| Width on the top | 32 |
| Thickness | 3.012 |
| Bottom of water-barrier | 0.43 |
| Top of water-barrier | 0.15 |
| Height of water-barrier | 1.2 |
| Distance between the water-barrier to the edge of the top surface | 0.3 |

According to the configurations, the water-filled traffic barriers have a height of 1.2 m and are placed 0.3 m away from the edge of the main beam.

### 3.2 Data Grouping

Based on our experiments, there are 10 cases in total. Four cases have a zero-degree angle of attack between the direction of the wind and the suspended bridge. A no-water barrier, one water barrier on the left, one water barrier on the right, and two water barriers are considered in these cases, respectively. Additionally, there are three cases with a negative angle of attack between the wind direction and the suspended bridge; the angles of attack are –5°, –10°, and –15°. Moreover, there are three cases with a positive angle of attack between the wind direction and the suspended bridge; the angles of attack are 5°, 10°, and 15°.

### 3.3 Data Processing Flow

As shown in Fig. 2, the data processing flow includes three main steps. In the pre-processing step, the data of the suspended bridge design drawings must be investigated. Then, the ANSYS software can be used to simplify the blueprint into a 2D model, and ANSYS ICEM can be used to generate the mesh. In the finite element simulation process, the boundary conditions are set based on measurements obtained from the literature, and the governing equations are solved to compute the flow field using Galerkin's Finite Element Method. To obtain a numerical solution to a differential equation, the domain is subdivided into finite elements. Trial piecewise functions approximate the function over each of these elements. Then, the flow field is exported to Tecplot and flow analysis is carried out. In the Post-processing step, the velocity and pressure are determined based on the computed flow field, the hemodynamic parameters are exported to Tecplot, and hemodynamic analysis is carried out.

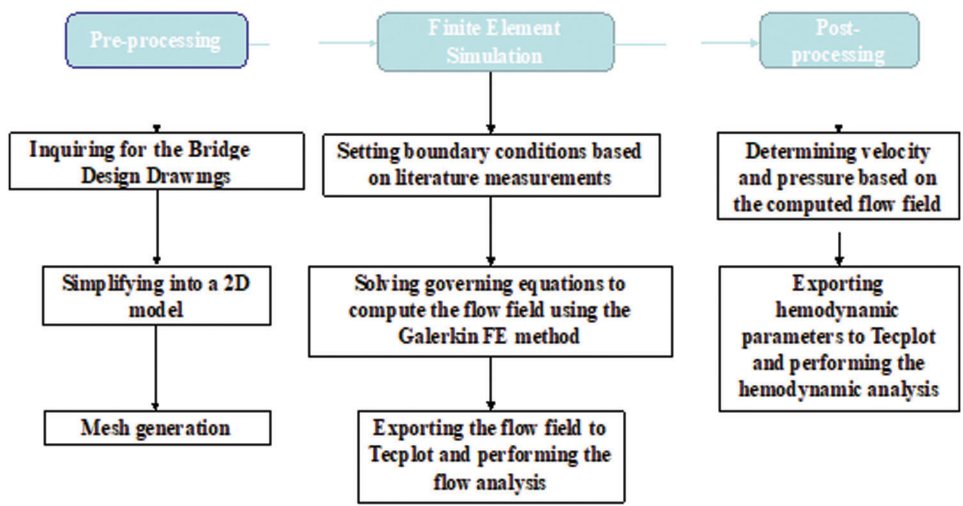

**Figure 2:** Data processing flow

## 4 Computational Fluid Dynamics Simulation

### 4.1 Governing Equations

The fluid flow is governed by the conservation laws in physics. In this study, aerodynamics analysis was carried out on the water-filled barriers of the suspended bridge using the conservation of mass equation and conservation of momentum equation (Navier–Strokes equation).

The incompressible flow obeys the mass conservation equation, and its mathematical expression can be written as follows:

$$\frac{\partial(u)}{\partial x} + \frac{\partial(v)}{\partial y} + \frac{\partial(w)}{\partial z} = 0 \tag{1}$$

We introduce the symbol of divergence, namely $\nabla \cdot a = div(a) = \partial a_x/\partial x + \partial a_y/\partial y + \partial a_z/\partial z$, and the expression in Eq. (1) becomes as follows:

$$\nabla \cdot v = 0 \tag{2}$$

The Conservation of Momentum obeys Newton's second law whereby the rate of change of the momentum of a microelement body is equal to the net external force that the body receives, as follows:

$$\frac{\partial v}{\partial t} + (v \cdot \nabla)v = -\frac{\nabla p}{\rho} + v\nabla^2 v \tag{3}$$

In Eq. (3), $\rho$ is the fluid density, v is the fluid velocity, $p$ is the pressure, $t$ is time.

### 4.2 Meshing

This study considered 10 groups of meshed two-dimensional models. In commercial CFD, meshing consumes a lot of time and energy. Moreover, the meshing quality is an important factor with regard to the computation speed and accuracy of the results. Here, ICEM was selected to divide the mesh. Notably, ICEM is a professional meshing, pre-processing, and post-processing software that interfaces with FLUENT and provides tools for geometry acquisition, mesh generation, mesh optimization, and post-processing to satisfy various complex geometric structures.

This study used ICEM to divide the model into quadrilateral and triangular surface meshes, as shown in Figs. 3 and 4. In each model, the mesh consisted of 313,870 elements and the minimum face area was 2.499765e–05 m².

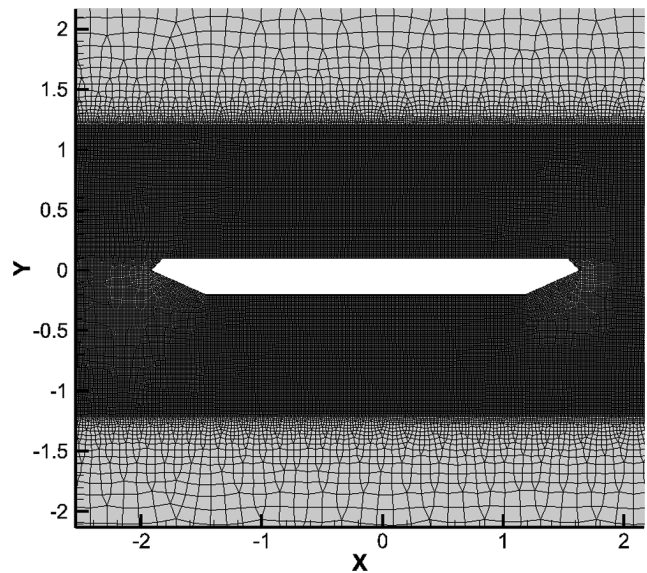

**Figure 3:** Mesh of suspension bridge without water barrier

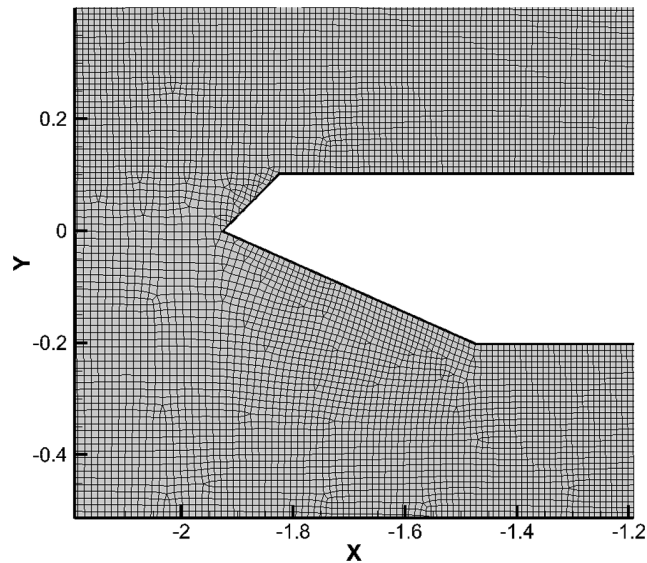

**Figure 4:** Mesh details of suspension bridge without water barrier

### 4.3 Discretization Schemes

The equation can be solved after the governing equation is discretized. This study used the PISO algorithm in the finite volume method to discretize the two-dimensional structure of the bridge–water striders.

The discretization of the computational domain consists of dividing the overall computational region into many subregions in the form of grids and transforming the above-mentioned basic governing equations into algebraic equations at each grid node. Computation technology is used to solve the numerical solutions of these equations and determine the approximate solutions between nodes using the interpolation method. When the mesh is sufficiently fine, the approximate solution of the discrete equation will approach the exact solution. The finite volume method is the most widely used and most mature method in CFD engineering. The computational domain is divided into a grid. Then, a non-repetitive governing volume is formed around each grid node. The fluid governing differential equation for each governing volume is integrated to form a set of discrete equations. The dependent variable $\varphi$ of each grid node is considered as the unknown quantity of the discrete equation system. The schemes used in discretization include the central difference scheme, first-order/second-order upwind scheme, and quick scheme.

The PISO algorithm is an optimized algorithm based on the close approximation of the pressure-velocity correction. Compared with the SIMPLE and SIMPLEC algorithms, the PISO algorithm requires slightly more computational time for each iteration but greatly reduces the number of iterations needed to achieve convergence. The processing flow of the PISO algorithm is shown in Fig. 5.

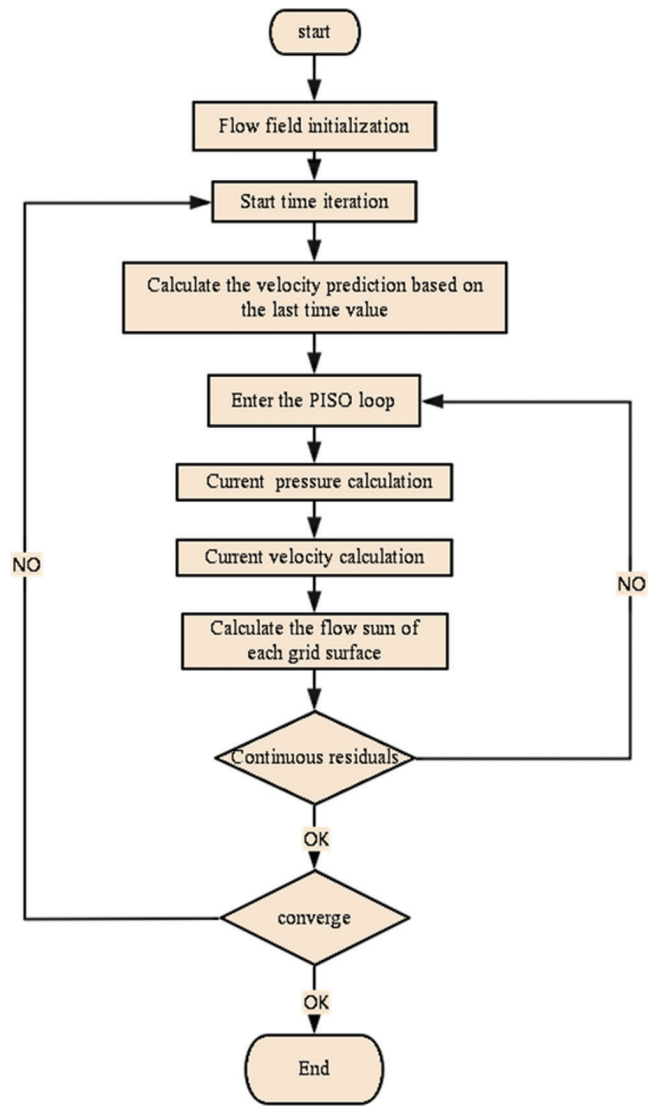

**Figure 5:** Processing flow of PISO algorithm

### 4.4 Turbulence Model

Computational fluid dynamics mainly include the Direct Numerical Simulation (DNS), Large Eddy Simulation (LES), and Reynolds Average N-S model (RANS) for solving turbulence.

The SST k-ω model in the RANS method was used in this study. The RANS method solves the N-S Eq. (1) of the overall mean (or time mean). All turbulence scales are simulated in the RANS method, which is currently the most widely used equation in industrial flow calculations. The RANS models provided by the CFD engineering software FLUENT mainly include the following.

The standard k-ε model includes a two-equation model and the default k-ε model. The coefficients are given by empirical formulas. The result is satisfactory for high Re turbulence, including for options such as viscous heat, buoyancy, and compressibility. The amount of calculation is moderate, more data are accumulated, and the accuracy is higher. Additionally, the model's convergence and calculation accuracy can satisfy the general engineering calculation requirements. However, the simulation is not suitable to complex flows with a large curvature and strong pressure gradient.

*4.5 Boundary Conditions*

In this study, the velocity inlet boundary condition and pressure outlet boundary condition were used to calculate the simplified two-dimensional model of the bridge. Additionally, a large area of space in which the wind flows was artificially defined. No-slip boundary conditions were adopted for both the bridge and the space wall surface. All walls had no-slip boundary conditions.

A brief introduction to these commonly used boundary conditions is presented below. Velocity-inlet refers to the inlet velocity and all scalar values that need to be calculated. This boundary condition is suitable for incompressible flow problems but not for compressible problems. Otherwise, the inlet boundary conditions will cause certain fluctuations in the total temperature or pressure at the inlet. Pressure-outlet refers to the static pressure of a given flow outlet. For exits with backflow, this boundary condition can converge more easily compared with the outflow boundary condition. However, this boundary condition can only be used to simulate subsonic flow.

For viscous flow problems, the default setting of FLUENT is the no-slip condition on the wall. When the wall has translational or rotational motion, the wall tangential velocity component can be specified, or the wall shear stress can be considered to simulate the wall slip.

# 5 Results and Discussion

*5.1 Mesh Sensitivity Analysis for Water Barrier in Different Positions*

This section discusses the CFD calculations that were carried out on the flow field of different bridge cross-sections under a wind incidence angle of 0°. These are the cross-sections with water-filled traffic barriers at both ends, one water-filled traffic barrier at the left end (near the entrance side), one water-filled traffic barrier at the right end (near the exit side), and no water-filled barriers. The inlet velocity was set to 80 m/s because the length scale of the simulation model was 1/10 of the actual length.

To more accurately simulate the characteristics of the flow field around the stationary cross-section and capture the vortex shedding characteristics, transient calculation was adopted in the CFD calculation. Because the value of the time step in the transient calculation may affect the accuracy of the results, after passing the irrelevance test and considering the calculation efficiency, the step size in this simulation was selected as 0.002 s.

In the mesh-based Finite Element Method, the quality of the mesh determines the success or failure of the analysis and the accuracy of the results. Research has shown that, when investigating complex geometries, finer meshes must be used in complex areas wherein large structural changes are expected. Although a simplified bridge model was used in this study, the mesh in the area near the bridge body was refined and optimized. The optimal mesh density was also selected to reduce the simulation time. Table 2 presents a grid sensitivity analysis, which compares the grid density adopted in this paper with the speed and pressure results obtained after the grid density increased. Five examples are considered to compare the calculation results before and after the mesh refinement. The results reveal that the mesh density used in this study is acceptable.

**Table 2:** Mesh sensitivity analysis

|  | Velocity error/% | Frequency error/% |
|---|---|---|
| Left side water barrier | 1.2% | 1.8% |
| Right side water barrier | 1.0% | 1.6% |
| Both side water barrier | 2.0% | 2.0% |
| Attack angle of −10° | 2.3% | 1.8% |
| Attack angle of 10° | 3.0% | 1.9% |

### 5.2 Position of Water Barrier Affecting Vortex Frequency

Based on the calculation result obtained with a time step of 0.002 s, the post-processing software CFD-Post was used to observe and analyze the vortex distribution and shedding in a stable period of the flow field when the original section was stationary. Fig. 6 shows the vorticity distribution and streamline diagram of the bridge cross-section without any water barriers.

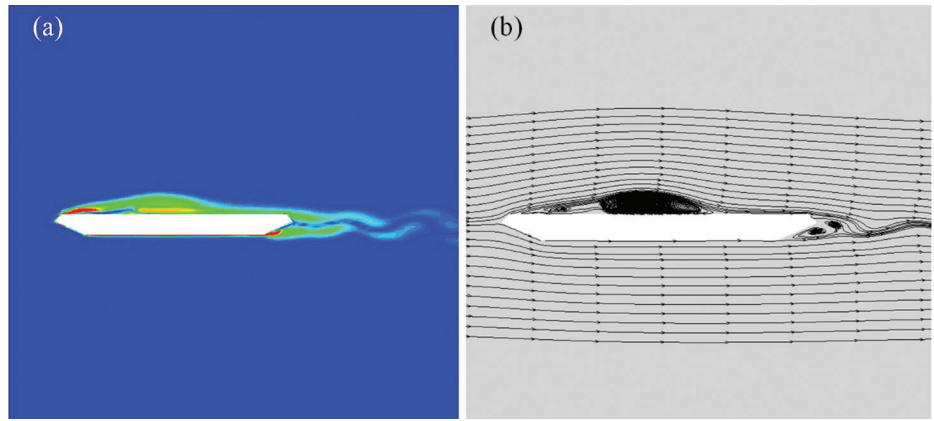

**Figure 6:** Flow field of cross-section without any water barriers: (a) vorticity distribution; (b) streamline diagram

Owing to the streamline shape of the cross-section without any water-filled traffic barriers, the airflow on the upper and lower surfaces of the cross-section almost touched the cross-section. A small degree of flow separation occurred only at the wind fairing, owing to the sudden change in the cross-sectional area of the flow channel. The flow separation near the wind fairing was very small. However, at the distal wind fairing, an "S" shaped vortex eventually formed in the wake flow area owing to the rapid change in the cross-sectional area of the lower surface of the flow channel.

There are two main methods for measuring the vortex shedding frequency: one is to measure the wind speed change at a certain point in the wake flow, and the other is to measure the lift force change in the cross-section. In this study, the lift force change was considered to measure the vortex shedding frequency. For the condition without water barriers, the Fourier transform spectrum of the lift force of the section is shown in Fig. 7. Under this condition, the vortex shedding frequency of the stationary flow was calculated as 7.35 Hz.

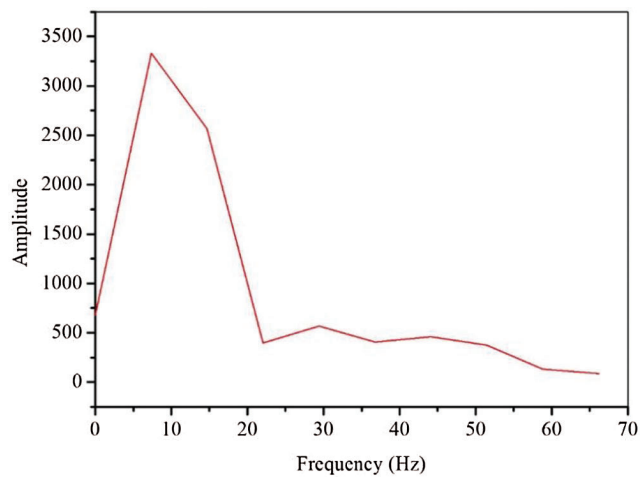

**Figure 7:** Spectrum of lift

When a water barrier is placed on the bridge, the flow field around the bridge cross-section changes accordingly. Fig. 8 shows the vorticity distribution of the cross-sections with water barriers on the left side, right side, and both sides, respectively.

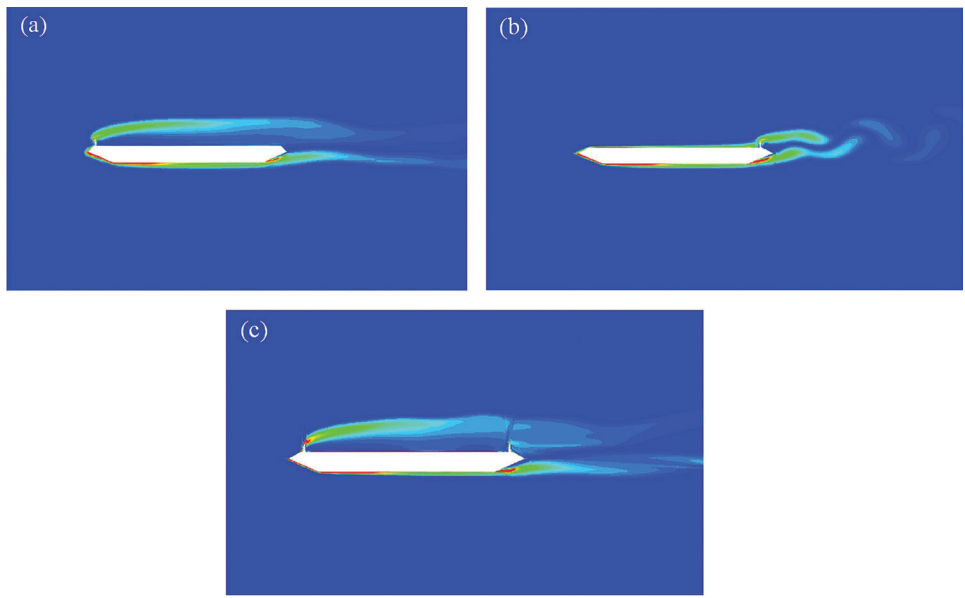

**Figure 8:** Vorticity distribution: (a) water barrier on left side; (b) water barrier on right side; (c) water barrier on both sides

For comparison to the no-water-barrier condition, Figs. 8 and 9 show the flow separation and non-negligible vortices caused by the water barriers on the surface of the section. These vortices continued to fall off the wake flow. When a water barrier existed at the near end, the airflow on the surface of the section produced a major vortex with large intensity and size after flowing through the water barrier. The strength of this vortex was greater than the vortex generated at the distal end of the lower surface, which made the vortex at the distal end of the lower surface dominant. Therefore, the wake vortices mainly depended on the surface vortices on the cross-section, and eventually fell off in an S-shaped manner.

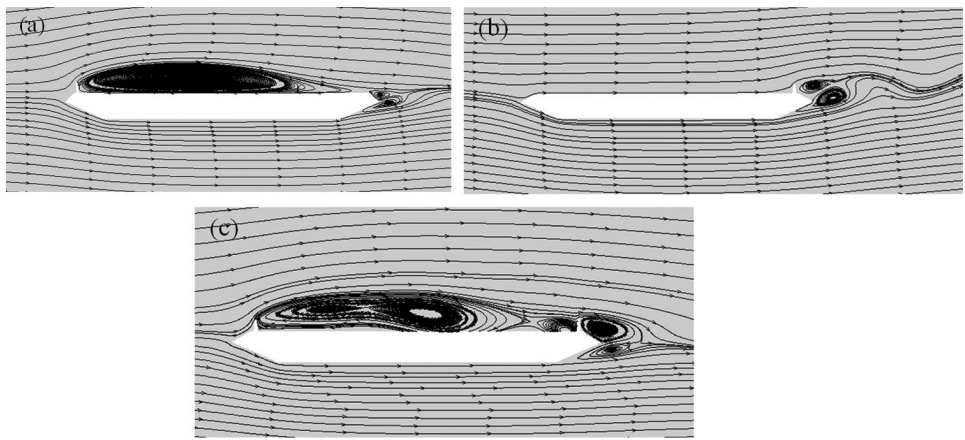

**Figure 9:** Streamline diagram: (a) water barrier on left side; (b) water barrier on right side; (c) water barrier on both sides

For the working conditions with a water barrier at the far end, the strength and size of the vortex generated by the airflow on the surface of the section after passing through the water barrier at the far end are equivalent to the strength and size of the wake vortex on the lower surface. A 2S-shaped wake vortex was formed, and this 2S form of wake vortex shedding can easily cause the vertical bending vortex vibration response of the bridge.

For the cross-sections with water barriers at both ends, the existence of water barriers at the distal end broke the large-scale main vortices generated by the airflow on the surface of the cross-section through the water barriers at the near end. This made the vortices around the section smaller and more numerous. The regularity of the cross-section flow field was destroyed, and the turbulence component in the flow field increased.

Fig. 10 shows the cross-section lift spectrum in the three simulations with water barriers at the near end, water barriers at the distal end, and water barriers at both ends. The calculation results obtained for the vortex shedding frequency of the two-static circumfluence (one around a water barrier at the near end and one around a water barrier at the far end) are 10.76 and 41.30 Hz, respectively. The vortex shedding frequency of the water barrier at the far end is close to the natural frequency of the bridge. Furthermore, the cross-section with water barriers at both ends broke the main vortex on the surface of the section owing to the water barrier at the far end. This resulted in increased turbulence in the flow field and more complexity in the aerodynamic pulsation, and various frequencies such as 7.26, 14.52, and 20.16 Hz were included.

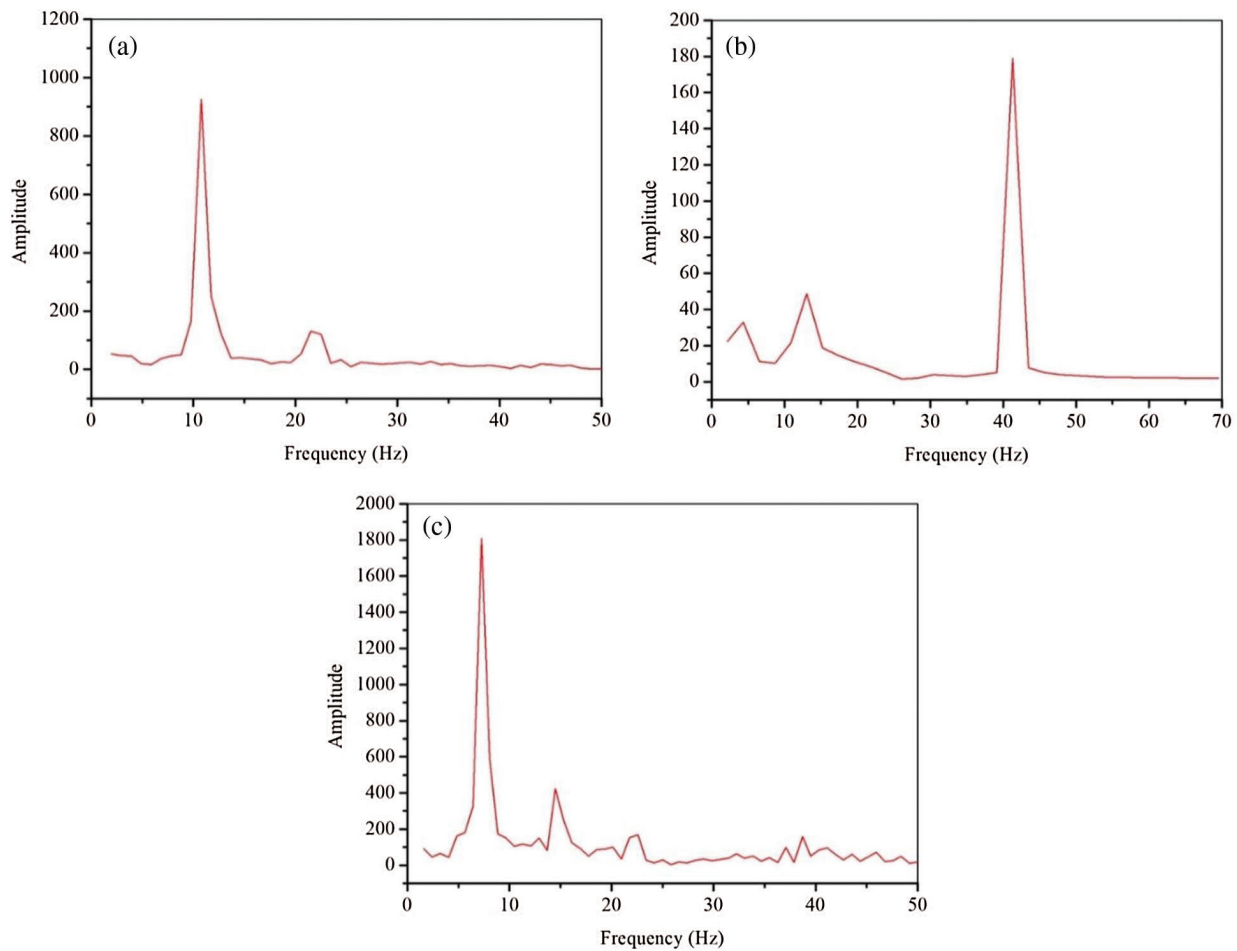

**Figure 10:** Spectrum of lift: (a) water barrier on left side; (b) water barrier on right side; (c) water barrier on both sides

### 5.3 Impact of Attack Angle on Bridge Vibration

Because the wind field of the bridge is strongly affected by the local topography and tends to generate incoming wind at an angle of attack greater than ±10°, it is necessary to investigate the frequency of vortex shedding at a high angle of attack. This section describes the CFD simulations that were carried out under water barrier conditions at both ends at six angles of attack of the bridge section: 15°, –10°, –5°, 5°, 10°, and 15°.

Fig. 11 shows the vorticity distribution at six angles of attack (–15°, –10°, –5°, 5°, 10°, and 15°) at the cross-section of the bridge with a water barrier at both ends.

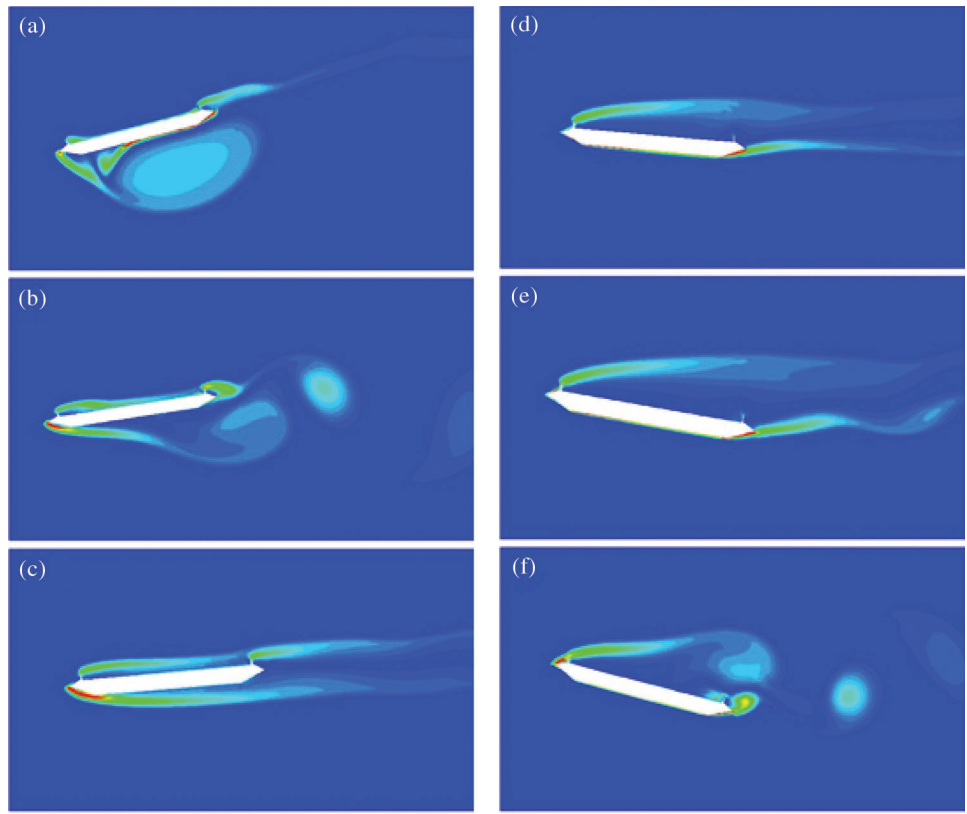

**Figure 11:** Vorticity distribution at different angles of attack: (a) attack angle of –15°; (b) attack angle of –10°; (c) attack angle of –5°; (d) attack angle of 5°; (e) attack angle of 10°; (f) attack angle of 15°

Under the action of a large negative angle of attack (–15°), the airflow was close to the surface of the cross-section. The near-end water barrier had a small effect on the airflow at the upper surface. The distal water barrier was directly exposed to the airflow. After the surface air flowed through the remote water barrier, flow separation occurred and a vortex was generated. On the lower surface of the section, the flow was completely separated owing to the large negative angle of attack, and reattachment did not occur on the lower surface of the section. Some vortices with a large range and strength continuously formed and moved toward the wake. In this case, the surface vortices were dominant.

Under the action of a medium negative angle of attack (–10°), the surface of the cross-section still flowed close to the cross-section until the upper surface air flowed through the remote water barrier and separated. On the lower surface of the section, the flow was completely separated because the negative angle of attack was still large, but the intensity and range of the generated vortices were less than those in the case wherein the angle of attack was –15°. At this time, in the wake area, the vortices generated on the upper and lower surfaces alternately fell off, while the vortices on the lower surface were still dominant.

Under the action of a small negative angle of attack (–5°), the airflow on the upper surface of the cross-section reattached soon after separation and continued to flow close to the cross-section until flow separation occurred again after passing through the far-end water barrier. On the lower surface of the cross-section, the airflow attached onto the lower surface of the cross-section, owing to the smaller negative angle of attack, and then separated at the far end of the wind fairing. Under this condition, the wake zone was mainly the vortex generated by the water barrier on the upper surface.

Under the action of a small positive angle of attack (5°), the upper surface airflow passed through the near-end water barrier to a certain degree of flow separation, and reattachment onto the upper surface did not occur. The water barrier at the far end broke the large-scale main vortex to a certain extent. There was no flow airflow separation on the lower surface up to the tip of the distal wind fairing. Then, the flow separation of the lower surface airflow occurred, and a vortex was generated on the upper surface of the wind fairing. However, because the angle of attack was small at this time, the strength of the vortex range was also small. At this time, the wake field was dominated by the main vortex generated on the upper surface.

Under the action of a medium positive angle of attack (10°), the airflow on the upper surface caused large flow separation after passing through the near-end water barrier, and reattachment did not occur on the upper surface. The presence of water barriers at the far end had almost no effect on the main vortex at the upper surface. The air flow at the lower surface separated at the tip of the distal nozzle. Compared with the –5° working condition, a vortex with a certain strength was generated. At this time, the two vortices in the wake field were alternately separated, but were still generated by the upper surface. The main vortex was dominant.

Under the action of a larger positive angle of attack (15°), the vortex intensity and range generated by the upper and lower surface air currents were equal. At this time, in the wake area, the vortex intensity and range alternately fell off in a 2S shape.

At the angles of attack of –10°, –5°, 5°, and 15°, the lift spectrum of the water barrier bridge section at both ends is as shown in Fig. 13. Under the above-mentioned working conditions, the vortex shedding frequencies of the static surrounding flow were calculated as 11.50, 23.72, 10.03, and 15.83 Hz, respectively. Because the –15° and 10° working conditions have more turbulent components in the flow field and more complex aerodynamic pulsations, they contain more frequency components, as shown in Fig. 12. The main frequency components of the attack angles of –15° and 10° are 7.15 and 7.82 Hz, respectively.

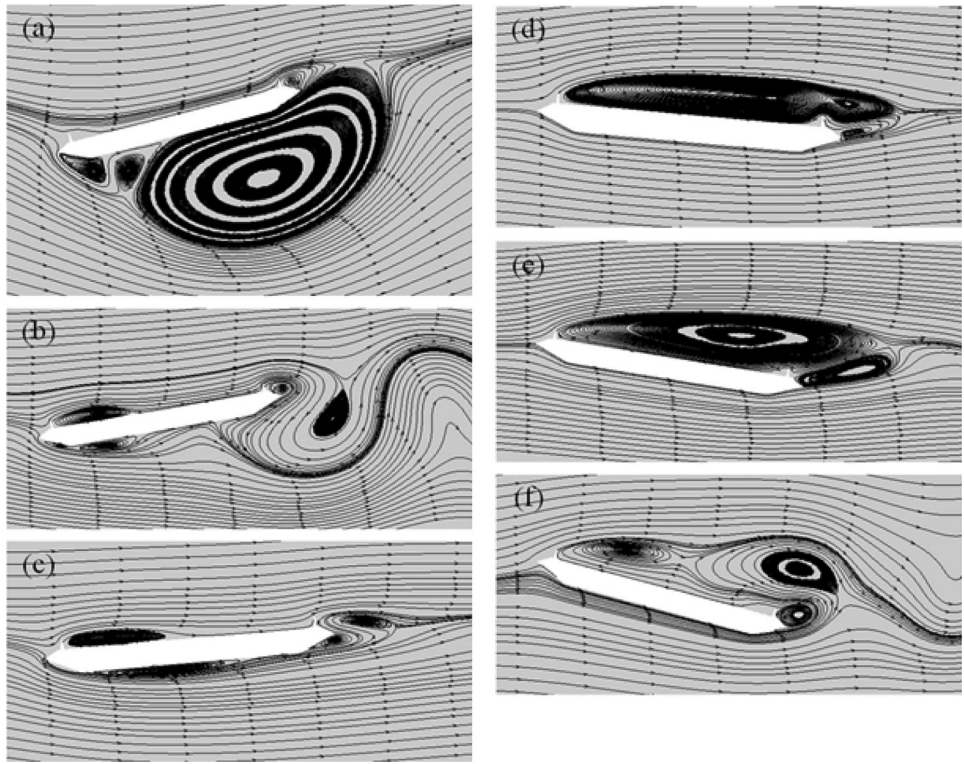

**Figure 12:** Streamline diagram of water barrier bridge section at both ends at (a) attack angle of −15°; (b) attack angle of −10°; (c) attack angle of −5°; (d) attack angle of 5°; (e) attack angle of 10°; (f) attack angle of 15°

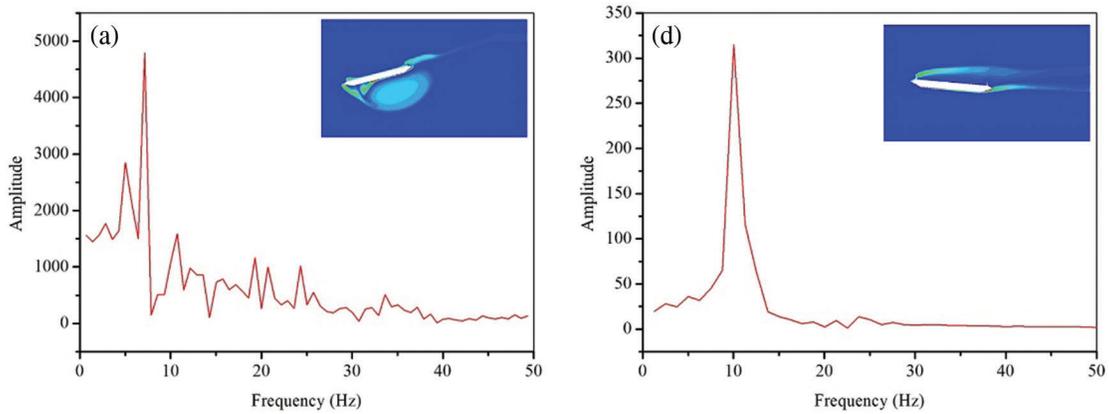

**Figure 13:** Continued

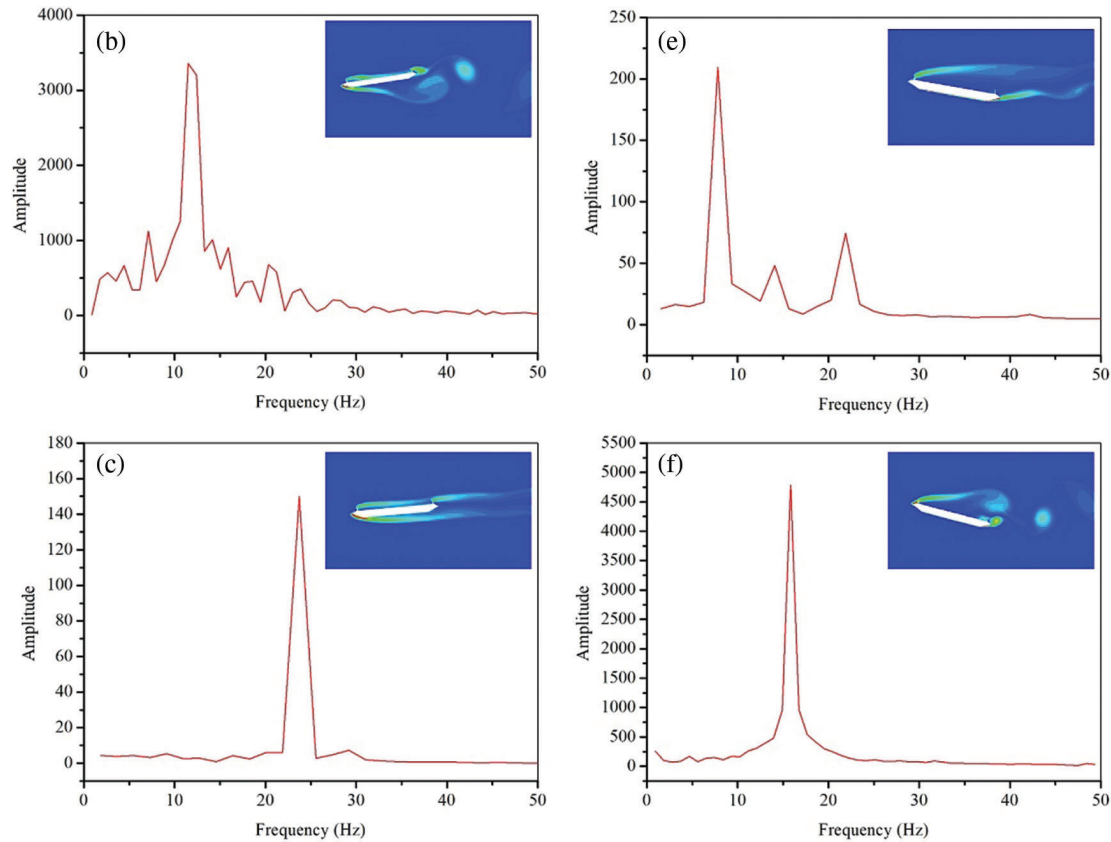

**Figure 13:** Spectrum of lift: (a) attack angle of –15°; (b) attack angle of –10°; (c) attack angle of –5°; (d) attack angle of 5°; (e) attack angle of 10°; (f) attack angle of 15°

## 6 Conclusions

This study first investigated how the position of water barriers on the wake vortex shedding affects the flow field around the bridge section under an attack angle of 0°. In the simulation considering a water barrier at the near end, the wake vortex mainly depended on the main vortex on the upper cross-section surface. In the simulation considering a water barrier at the distal end, the strength and size of the vortices on the upper and lower surfaces were equal, and formed a 2S-shaped wake vortex. In the simulation with water barriers at both ends, the number of vortices in the flow field around the bridge cross-section increased and contained multiple frequencies. Notably, the vortex shedding frequency is similar to the resonance frequency of the sample suspended bridge under the condition of water barriers existing at the far end. Moreover, the vortex shedding of the 2S-shaped wake can easily cause the vertical vortex-bending response of the bridge.

This study also investigated how different incoming angles of attack affect the flow field around the bridge cross-section and the vortex detachment frequency. In the simulations with negative incoming angles of attack, when the angle was small (–5°), the circumferential vortex flow field and wake vortex detachment frequency were mainly affected by the flow separation caused by the water barrier at the far end, and the simulated vortex detachment frequency was similar to the resonance frequency of the sample suspended bridge. As the negative angle of attack increased to –10° and –15°, the flow on the lower surface completely separated, and the circumferential vortex flow field and wake vortex detachment frequency were mainly affected by the main vortex formed under the lower surface. In the simulations with positive angles of attack, when the angle was small (5°), the main vortex generated at the upper surface was dominant. When the positive angle of attack gradually increased to 10° and 15°, the intensity and size of the vortex generated at the tip of the distal wind fairing under the lower surface of the bridge gradually increased, which made the two vortices of the wake flow field fall off in an S shape manner rather than a 2S shape manner.

There are a lot of areas for improvement that can be addressed in future work. For example, a three-dimensional model with different variables can be used to improve the simulation accuracy. The effect of different weather conditions on the vortex vibrations of the bridge should also be analyzed. More simulations under different circumstances, such as different weather conditions, barriers, and different types of aerodynamic structures, should be investigated.

**Acknowledgement:** Thanks are due to Dongjie Niu for assistance with valuable discussion.

**Funding Statement:** The authors received no specific funding for this study.

**Conflicts of Interest:** The authors declare that they have no conflicts of interest to report regarding the present study.